# Application of Ultraviolet Fluorescence in Wafer Surface Cleaning Analysis

Zhongwei Pan*
*Department of Electrical Engineering, the City College of the City University of New York, New York, NY 10031*

Richard Chin
*Computer Network Services Inc., 100 Ford Road, Denville, NJ 07834*

Abstract

Ultraviolet fluorescence (UVF) is introduced as a novel and versatile method for the determination of the contamination of metal impurities on silicon wafer surfaces. The results given demonstrate the usefulness of UVF for contamination control in silicon wafer processing. The experiments show some main characteristic peaks such as Fe, Cu, Ni and Na at room temperature. Compared with other surface sensitive techniques, the main advantages of UVF are low detection limits, simultaneous multi-element analysis and high sensitivity (up to $10^9$ Fe/cm$^2$).

## INTRODUCTION

Every wafer-processing step is a potential source of contamination, which may lead to defect formation and device failure. Cleaning of wafers must take place after each processing step and before each high-temperature operation[1,2]. With shrinking device dimensions and thickness of insulating layers down to 10 nm, it becomes imperative to avoid metal contamination to an unprecedented level. Therefore, a comprehensive control and regular monitoring technique of contamination levels is an important prerequisite for obtaining high, stable device yields and good device reliabilities.

Some metallic impurities such as Fe, Cu, Ni and Na may be incorporated into silicon wafers during some processing steps, such as thermal oxidation, reactive ion etching and ion implantation[3]. Fe and Na have been considered as the most harmful impurities to cause the degradation of silicon-based VLSI devices. The traditional surface sensitive techniques, like Auger electron spectroscopy and x-ray photoelectron spectroscopy (XPS) are hampered by too high detection limits of about $10^{13}$ atoms/cm$^2$. Recently, it is demonstrated that metal contamination on silicon-based VLSI devices leads to degradation of electrical parameters below levels of $10^{10}$ atoms/cm$^{-2}$[4]. The MOS C-t, surface photovoltage (SPV), deep-level transient spectroscopy (DLTS) and total reflection X-ray fluorescence (TXRF) measurement methods are no sensitive to Na contamination. So far, there have been not yet any available methods to simultaneously detect Fe and Na contamination. In this paper, we first report a new technique — ultraviolet fluorescence (UVF) to be used as a fast, practically preparation-free, nondestructive detection method for evaluating the performance of cleanliness of VLSI wafers in real processes.

## EXPERIMENTAL DETAILS

The samples are measured by the recording spectro-fluorophotometer (F 3010-type). The light source is xenon lamp (150 w). The measurement system is fully controlled by a computer, and wafers up to 6 inch in diameter can be investigated. The measured area on a wafer can be varied between about 1×1 and 5×5 cm$^2$. The grating constant of the excitation and fluorescent monochromator is 900/mm. Their measurement accuracy can reach below 2 nm for wavelength. The injected beam of UV light only permeates into the silicon surface about 30 nm. All measurements were carried out on boron-implanted Czochralski (Cz) and floating-zone (FZ) wafers (3-6 inch in diameter). The boron concentration of the investigated samples covered a range of about $10^4$-$10^6$ cm$^{-3}$, but mostly samples with 1-2×$10^{15}$ cm$^{-3}$ were used. There was no preparation of the samples before the measurement.

* Electronic mail: zhwei@ee-mail.engr.ccny.cuny.edu

## RESULTS AND DISSCUSIONS

In the UVF measurement, we obtain the characteristic peaks of trace impurities Fe, Ni, Cu and Na in the range of 300-400 nm at room temperature when the excitation wavelength is in the range of 220-300 nm. These peak wavelengths are as follows: 371 nm, 383 nm and 398 nm of Fe, 330 nm of Na, 325nm and 378 nm of Cu, 341nm and 352 nm of Ni, respectively. Figure 1 shows the UVF spectrum of Fe and Na contamination, the specimens are from three different venders. Since pure silicon is not a fluorescence material in the range of 300-400 nm, it has no characteristic peak as shown in curve (a). Figure 2 shows the UVF spectrum of the failure VLSI chips.

As shown in Figure 3 the UVF peak intensities ($I_f$) are proportional to the product of the concentration (Ci) and fluorescent efficiency ($\Phi_i$) of the metal impurities. They obey Stokes law and can be written as Equation (1) when $(C_i/C_{mi}) << 1$.

$$I_f = \frac{A\Phi_i C_i}{C_{mi}} \qquad (1)$$

where A is constant, $C_{mi}$ is the concentration of mono-crystalline silicon. hence, the cleanliness level of silicon wafers can be evaluated according to above spectral peak intensities.

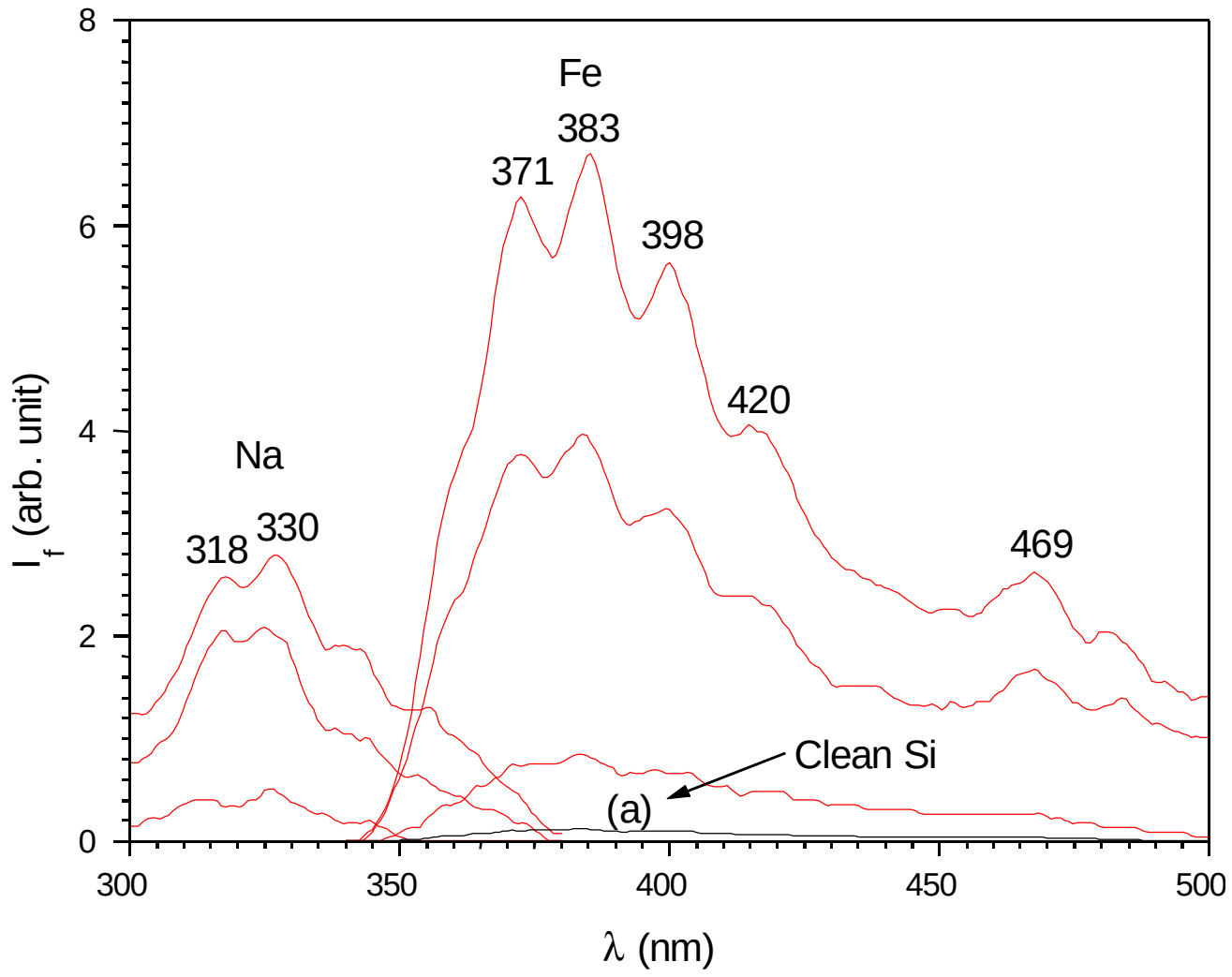

Fig. 1 UVF spectra of Fe and Na contamination of cleaning wafer after thermal oxidation. Curve (a) is the UVF spectrum before thermal oxidation.

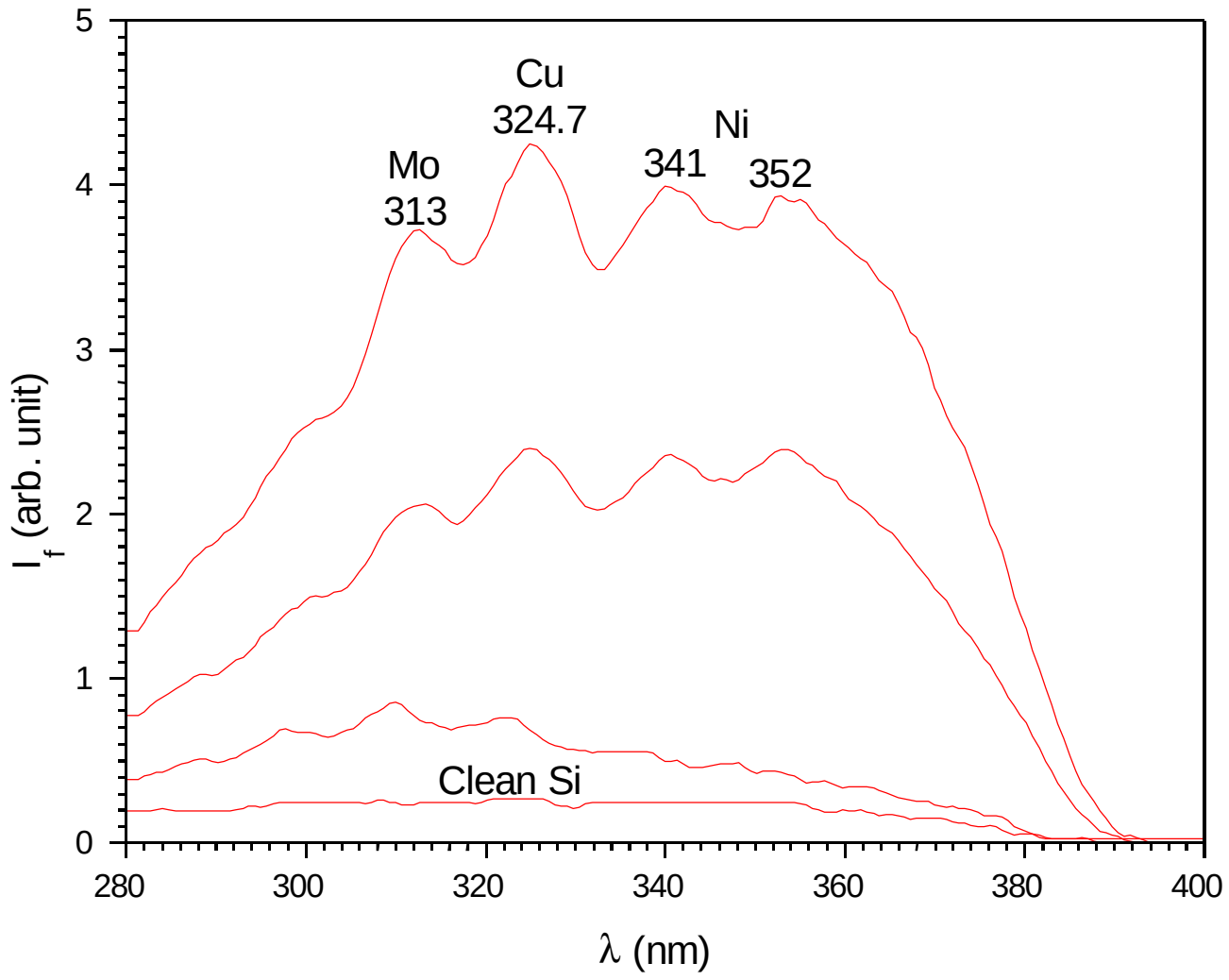

Fig. 2 UVF spectra of Cu, Ni and Mo impurities near silicon surface of failure VLSI chips.

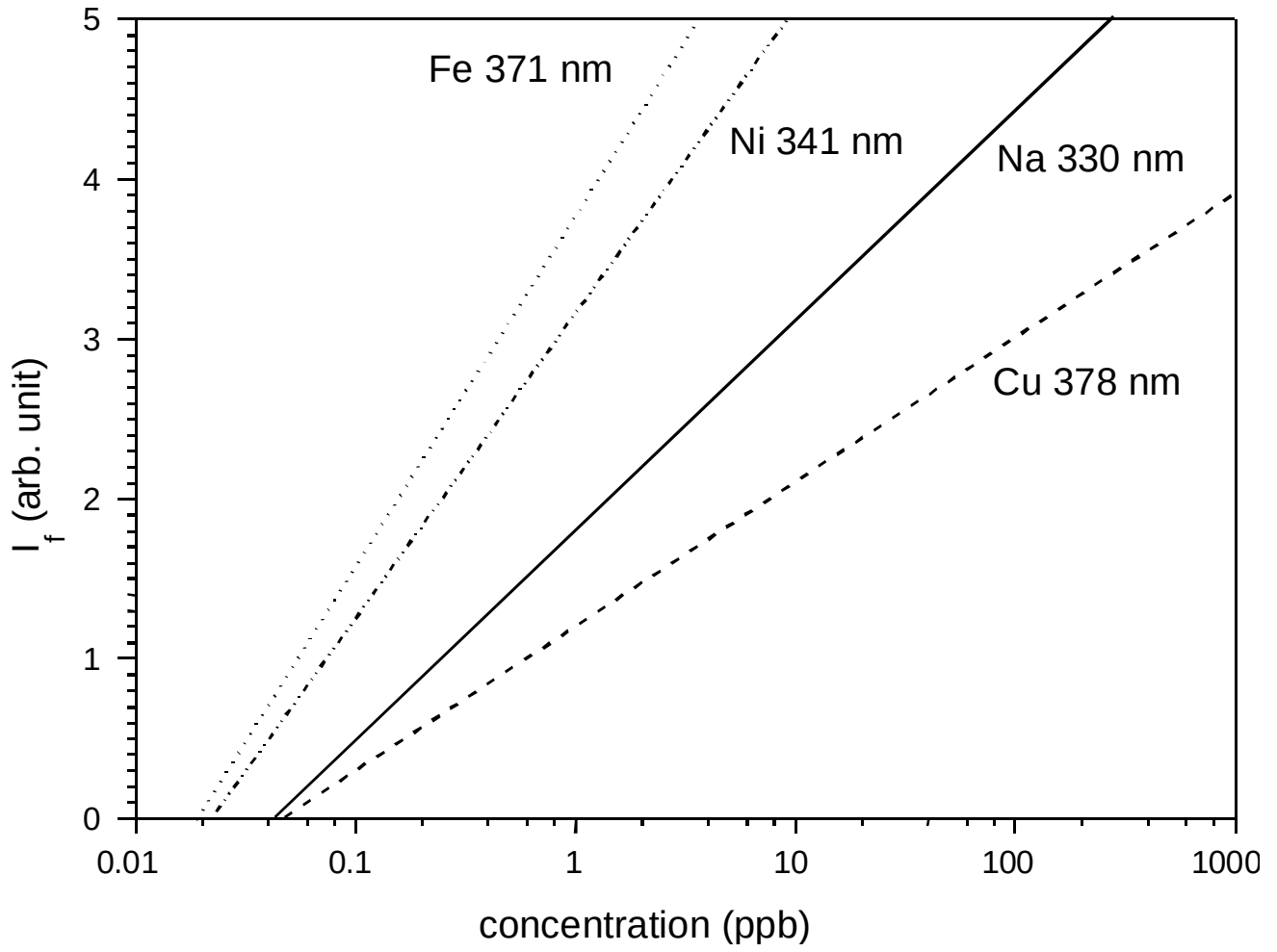

Fig. 3 shows fluorescence peak intensity vs the content of impurities of silicon-polished surfaces.

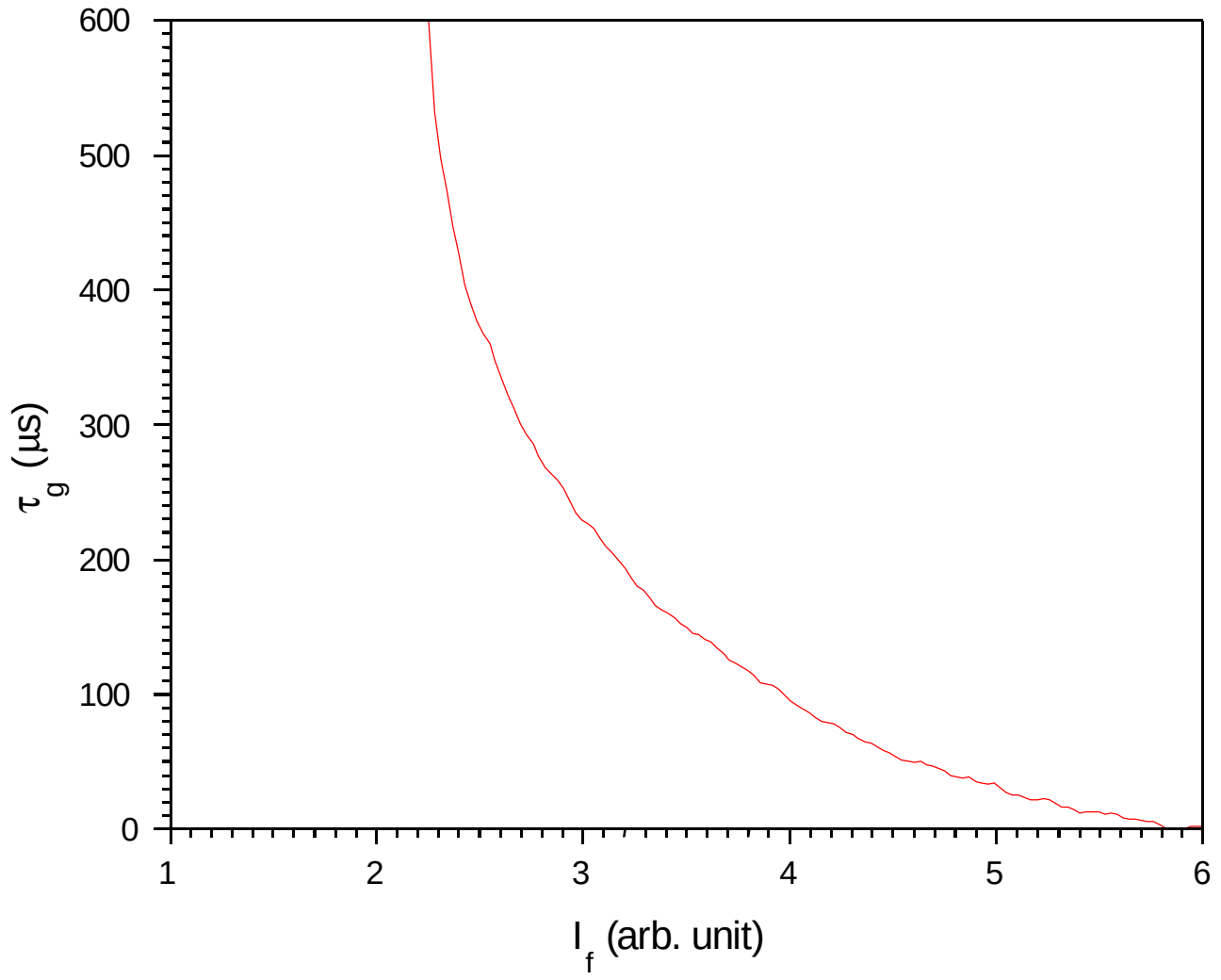

Fig. 4 the lifetime $\tau_g$ vs the contamination of Fe impurities on the thermal oxide wafers.

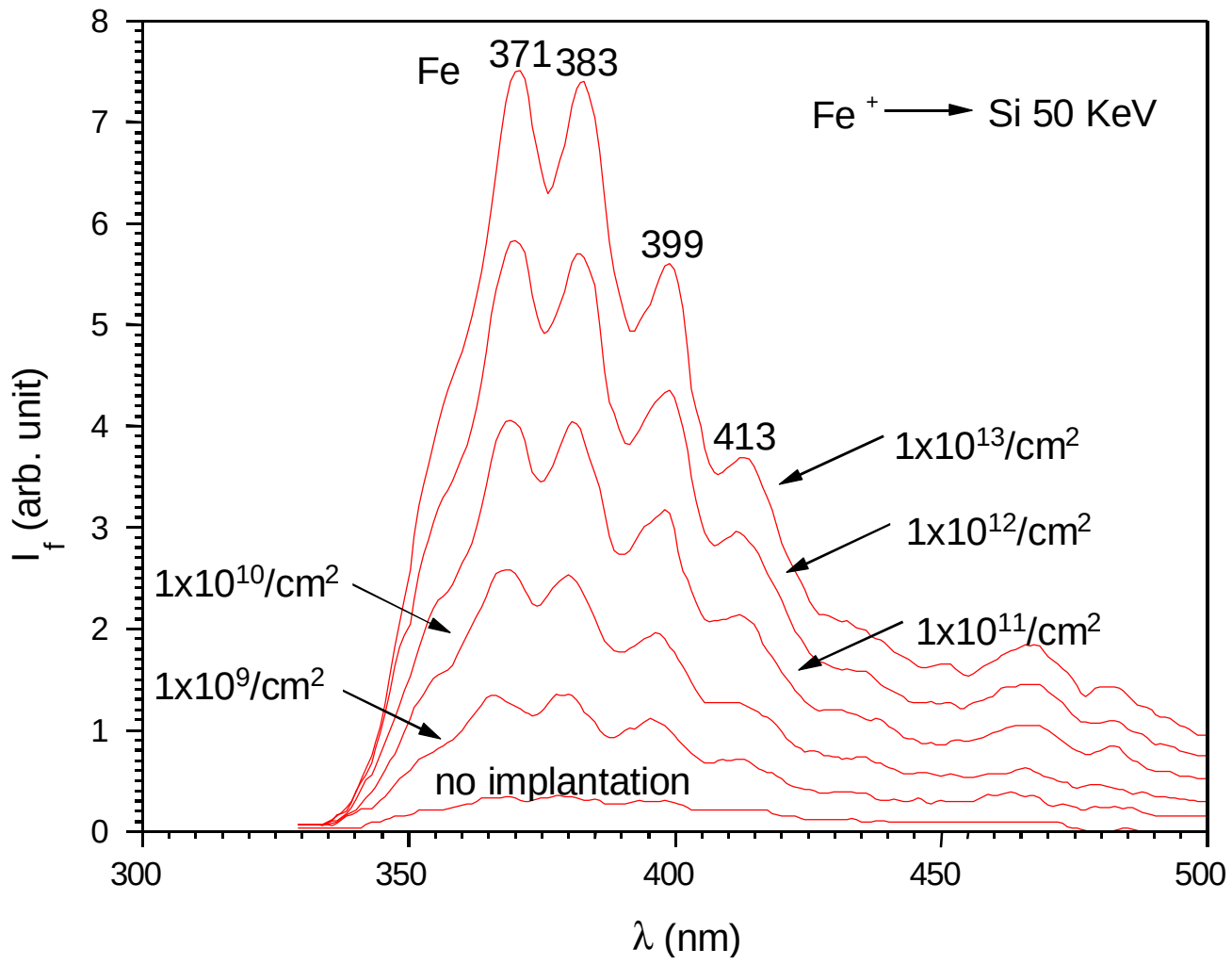

Fig. 5 the UVF spectra of the Fe ion implanted wafers which are annealed at 140 °C for 120 mins.

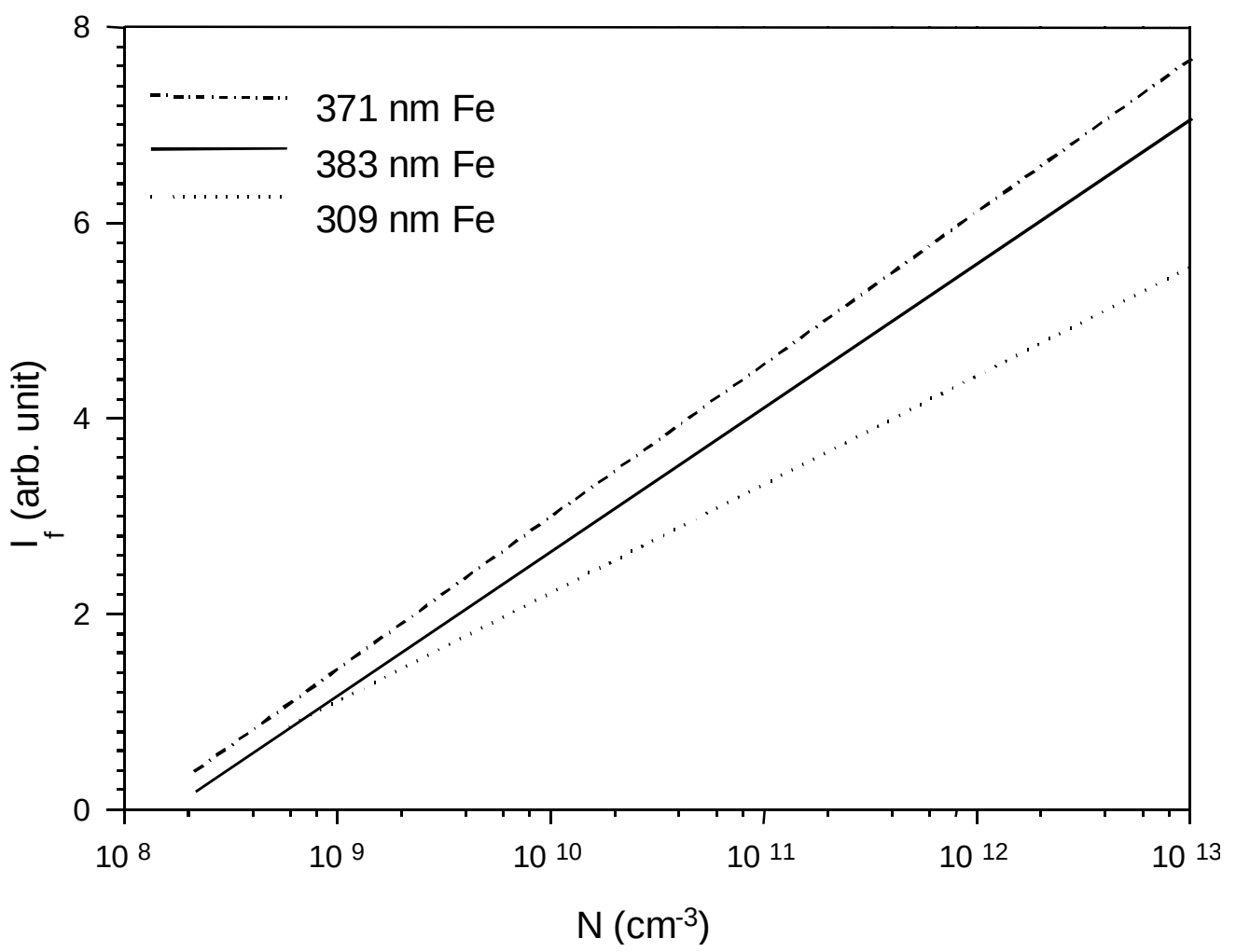

Fig. 6 shows the relationship of Fe fluorescence peak intensity and its ion implanted dose.

In Figure 4, the minority carrier generation lifetime $\tau_g$, which is measured by the MOS C-t method, is dropped rapidly as increase of Fe impurities near silicon surfaces under oxide film (about 600 Å thickness). In order to obtain the Fe-standard wafers, n-Si (100) 6 Ωcm wafers are implanted at 50 keV with dose: $1\times10^{9}$, $1\times10^{10}$, $1\times10^{11}$, $1\times10^{12}$ and $1\times10^{13}$ Fe/cm$^2$. Their corresponding Fe concentrations on the wafer surface are $1.3\times10^{13}$, $1.3\times10^{14}$, $1.3\times10^{15}$, $1.3\times10^{16}$, and $1.3\times10^{17}$ Fe/cm$^3$. Figure 5 shows the UVF spectra of Fe on the Fe-implanted wafers which were annealed at 140 °C for 120 min and Figure 6 shows the relationship between Fe fluorescence intensity and its ion implanted dose.

Finally, the performance of silicon wafers cleanliness (θ) can be expressed as:

$$\theta = \sum_i \theta_i = \sum_i \left( \frac{I_i}{I_{mi}} - 1 \right)^{-1} \qquad (2)$$

( i: Fe, Ni, Cu, Mo, Na, ......)

where $I_{mi}$ and $I_i$ are fluorescence intensity of metal impurities on the standard samples and on the detected wafers, respectively. Obviously, the smaller the content of metal impurities, and the less the type of the metal impurities, the higher the performance of cleanliness of wafer surface. Further investigation is in progress.

## CONCLUSIONS

Ultraviolet fluorescence measurement is a sensitive method for detecting the contamination of Fe, Cu, Ni and Na near the surface of silicon wafers, the high sensitivity up to $10^{13}$-$10^{14}$ Fe/cm$^3$ ($10^{8}$-$10^{9}$ Fe/cm$^2$) can be achieved. The results show that this method is simple and effective to monitor and evaluate the cleanliness performance of VLSI wafers in real processes.

## ACKNOWLEDGMENTS

The authors would like to thank the support of the New York State Science and Technology Foundation through its CUNY Center for Advanced Technology on Photonic Materials and Applications. Financial support from Reveo Inc. is gratefully acknowledged.